%% file: paper.tex
\documentclass[pra, twocolumn, showpacs]{revtex4-1}

\usepackage{amsmath}
\usepackage{color}
\usepackage{hyperref}
\usepackage{MissingT}
\usepackage{graphicx}

\newcommand{\emphdef}[1]{\emph{#1}}
\newcommand{\refeq}[1]  {Eqn \ref{#1}}
\newcommand{\reffig}[1] {Fig \ref{#1}}
\newcommand{\refsec}[1] {Section \ref{#1}}
\newcommand{\reftab}[1] {Table \ref{#1}}

\begin{document}

\title{Simulations of Shor's Algorithm using \\ Matrix Product States}

\author{D. S. Wang}
\email{dswang@physics.unimelb.edu.au}

\affiliation{Centre for Quantum Computation and Communication Technology, \\
School of Physics, The University of Melbourne, Parkville, Victoria 3010, Australia}

\author{C. D. Hill}
\email{cdhill@unimelb.edu.au}

\affiliation{Centre for Quantum Computation and Communication Technology, \\
School of Physics, The University of Melbourne, Parkville, Victoria 3010, Australia}

\author{L. C. L. Hollenberg}
\email{lloydch@unimelb.edu.au}

\affiliation{Centre for Quantum Computation and Communication Technology, \\
School of Physics, The University of Melbourne, Parkville, Victoria 3010, Australia}



\begin{abstract}
We show that under the matrix product state formalism the states
produced in Shor's algorithm can be represented using
$O(\max(4lr^2, 2^{2l}))$ space, where $l$ is the number of bits in the
number to factorise, and $r$ is the order and the solution to the
related order-finding problem.  The reduction in space compared to an
amplitude formalism approach is significant, allowing simulations as
large as $42$ qubits to be run on a single processor with $32$GB RAM.
This approach is readily adapted to a distributed memory environment,
and we have simulated a $45$ qubit case using $8$ cores with $16$GB
RAM in approximately one hour.
\end{abstract}

\maketitle

\section{Introduction}

Running large-scale quantum algorithms on a quantum computer is still
a distant goal.  The qubit \cite{mike+ike} is a two-level system which
can be in superpositions of both states, and is the fundamental unit
of information for quantum computation.  Not only is creating large
numbers of qubits with precise control difficult, so too is preserving
the delicate entanglement.  Thus, classical simulation is essential to
study the tolerance of quantum algorithms against qubit decoherence
and control errors.

Shor's algorithm for prime number factorisation \cite{shor94,shor95pf}
is a well known quantum algorithm, and has been simulated extensively
in the amplitude formalism \cite{oben98,niwa02,taba09,raed07,juli10}.
The computational resources grow exponentially with system size and
the simulation results are easily verified, making it an ideal test
candidate.  Amplitude formalism simulators have simulated systems with
as many as $42$ qubits on massive supercomputers \cite{raed07,juli10},
using highly parallelised software.  These calculations represent the
current limit using the amplitude formalism.

Here we consider a different approach to classical quantum algorithm
simulation.  The \emphdef{matrix product state} (MPS) formalism
\cite{vida03} is a specialised form of tensor network where the
interacting bodies are confined to one dimension.  Briefly, each
quantum state is represented as a product of matrices.  The matrix
dimensions depend on the entanglement in the system, and exceptional
cases may lead to an efficient simulation.  Fortunately for quantum
computing, this is not the case for Shor's algorithm.  Nonetheless,
MPS can be useful for circuit simulation.

In this paper, we shall show that the space requirements for running
Shor's algorithm using a MPS simulator is $O(\max(4l r^2, 2^{2l}))$,
where $l$ is the number of bits in the number $N$ to factorise, and
$r$ is the order of the related order-finding problem.  The space
reduction compared to the $O(2^{3l})$ of amplitude formalism
simulators allows us to simulate a $42$ qubit case on a single
processor with $32$GB RAM in a matter of hours.  We discuss
optimisation techniques that arise from circuit.  We also reason that
MPS simulators are well suited to parallelisation, and demonstrate
near ideal scaling using MPI.  Our largest simulation instance
consisted of $45$ qubits, using $8$ processor cores with $16$GB RAM in
approximately one hour (real time).

This paper is organised as follows.
\refsec{sec:shor} is a review of Shor's algorithm.
\refsec{sec:mps} is a review of MPS.
\refsec{sec:analysis} analyses the space performance of Shor's
algorithm when stored as a MPS.
\refsec{sec:bench1} describes our implementation in detail, and
tabulates single process benchmarks.
\refsec{sec:bench2} discusses a multi-processor implementation, and
benchmarks indicate good scaling behaviour.


\input{chapters/shor}
\input{chapters/mps}
\input{chapters/analysis}
\input{chapters/bench1}
\input{chapters/table1}
\input{chapters/bench2}
\input{chapters/table2}

\section{Conclusion}

We have shown that, in order to factorise an $l$-bit number $N$ with
order $r$, a MPS approach reduces the space complexity from
$O(2^{3l})$ to $O(\max(4lr^2, 2^{2l}))$ in the case of dense storage.
This is a significant improvement as $r$ is typically much smaller
than $N$, allowing one to simulate circuits with as many as $36$
qubits on a standard laptops.  Our largest simulation on a single
processor of $42$ qubits required less than $32$GB RAM.

We have also ported our simulations to use MPI.  We have simulated
circuits with as many as $45$ qubits over $8$ CPU cores.  Since MPS is
built around matrix products, the distributed matrix approach to
parallelisation understandably works well.  Our benchmarks show near
$1/n_{\mathrm{proc}}$ scaling.

One can extend these simulations, for example, by implementing the
circuits for the controlled-unitaries, and by incorporating
decoherence.  Other quantum circuits with the same structure as the
order-finding circuit are likely to benefit from an MPS approach, and
many of the optimisations described here would be applicable to these
and other MPS simulations.

This research was conducted by the Australian Research Council Centre
of Excellence for Quantum Computation and Communication Technology
(project number CE110001027).  This research was supported in part by
the U.S. Army Research Office (W911NF-08-1-0527), the US National
Security Agency and the Albert Shimmins Memorial Fund.

\bibliography{paper}
\bibliographystyle{elsarticle-num}

\end{document}

%% file: chapters/shor.tex
\section{Review of Shor's Algorithm}
\label{sec:shor}

We shall give a brief review of Shor's algorithm
\cite{shor94,shor95pf}.  Given an $l$-bit number $N = p \times q$,
where the integers $p$ and $q$ are co-prime, the task is to determine
the factors $p$ and $q$.  This is canonically achieved by reduction to
a related \emphdef{order-finding} problem: for a chosen integer $x$,
where $1 < x < N$, find the \emphdef{order} $r$ of $x$ modulo $N$.
The order $r$ is the smallest positive integer satisfying
\begin{align}
x^r \equiv 1 \pmod N.
\label{eq:order}
\end{align}

\begin{figure}
\centering
\includegraphics{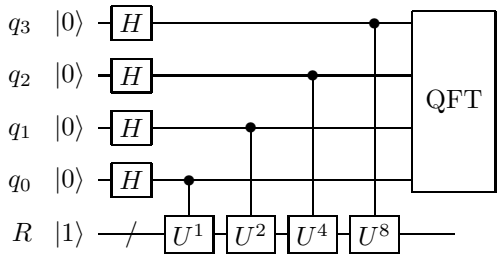}
\caption
{Schematic of the order-finding circuit which efficiently determines
the order $r$.  To factorise an $l$-bit number $N$, one uses an upper
register with $2l$ qubits, a lower register $R$ with $l$ qubits, and
$U\ket{a} = \ket{ax \bmod{N}}$ for some integer $1 < x < N$.
$\mathrm{QFT}$ is the quantum Fourier transform.}
\label{fig:order-finding schematic}
\end{figure}

\noindent
It can be shown that for even valued $r$ and
$x^{r/2} \neq -1 \pmod N$, one can find the factors of $N$ by
computing the greatest common denominator $\gcd(x^{r/2} \pm 1, N)$.
This can be performed efficiently by Euclid's algorithm.  The quantum
order-finding circuit (\reffig{fig:order-finding schematic}) that lies
at the heart of Shor's algorithm allows a quantum computer to
\emphdef{efficiently} determine the order $r$, using qubits and
circuitry growing only polynomially with $l$.  It consists of $3l$
qubits divided into two registers: the upper $2l$ qubits are
initialised in $\ket{0}$, and the remaining $l$ qubits are initialised
in $\ket{1}$.  In the figure, $H = (\sigma_x+\sigma_z)/\sqrt{2}$ is
the \emphdef{Hadamard} operation, where $\sigma_x$ and $\sigma_z$ are
the Pauli matrices, the unitary $U$ performs
\begin{align}
U \ket{a} = \ket{a x \mod N},
\end{align}

\noindent
and $\mathrm{QFT}$ denotes the \emphdef{quantum Fourier transform}
\cite{copp94}.  Neglecting normalisation constants, one can show that
the state created after the controlled-$U$s have been applied is
\begin{align}
\ket{\psi} = \sum_{i=0}^{2^{2l}-1} \ket{i} \ket{x^i \mod N}.
\end{align}

\noindent

We shall always suppress the Kronecker or tensor product between two
kets (i.e. $\ket{a} \ket{b} \equiv \ket{a} \otimes \ket{b}$).
Substituting in \refeq{eq:order} allows one to rewrite this as
\begin{align}
\ket{\psi}
= \sum_{i=0}^{r-1}
  \paren{\sum_{j=0} \ket{jr+i}} \ket{x^i \mod N}.
\label{eq:simplified state}
\end{align}

After applying all of the controlled-$U$s, one can measure and discard
the lower register, leaving a system containing only $2l$ qubits.  The
QFT produces a probability distribution in the upper register with
high probability densities in states around integer multiples of
$2^{2l}/r$.  The results of a few measurements from this probability
distribution can be classically processed into $r$ via the continued
fractions algorithm.  Since it is possible to construct circuits to
efficiently perform all powers of $U$ \cite{zalk98,beau03} and the QFT
\cite{copp94}, it is therefore possible for a quantum computer to
efficiently determine the order $r$ and hence factorise large numbers.

%% file: chapters/mps.tex
\section{Review of Matrix Product States}
\label{sec:mps}

Here we briefly review the MPS formalism \cite{vida03}.  The reader
may be familiar with \emphdef{amplitude formalism} state vector
simulators, where one stores the state as a collection of complex
coefficient-index pairs.  Typically this is implemented as a complex
number along a contiguous array.  This approach requires space
governed only by the size of the system, regardless of the state
stored.  The MPS formalism is a different way to store the state,
where the space required grows as the state becomes entangled, by
using the ability to write a general quantum state $\ket{\psi}$ as
\begin{align}
\ket{\psi}
= \sum_{a b c \cdots}
  \paren{A_a B_b C_c \cdots}
  \ket{a}\ket{b}\ket{c}\cdots.
\label{eq:general mps state}
\end{align}

\noindent
Here $\ket{a}$, $\ket{b}$, $\ket{c}$, etc. are the bases of the
subsystems of the state, and $A_a$, $B_b$, $C_c$, etc. are matrices
associated with the respective subsystems.  That is, the amplitude
$\alpha_{abc \cdots}$ of the state is stored as the matrix product
$A_a B_b C_c \cdots$.  The indices $a$, $b$, $c$, etc.  span the
dimensions of their respective subsystems (e.g. for a qubit, the index
can be zero or one), allowing for straightforward generalisations to
$d$-level subsystems (i.e. \emphdef{qudits}).  Indeed, any two or more
neighbouring qudits in a MPS---that is, their matrices are adjacent in
the product---can be combined into a single higher-dimensional qudit
simply by the evaluating the different matrix products and redefining
the subsystems (e.g. $D_d \equiv D_{ab} \equiv A_a B_b$).

A single-qudit gate $U_1$ is applied by mapping $U_1$ between the
corresponding elements in the matrices associated with that qudit
only.
Many-qudit gates (assuming all involved qudits form a contiguous
sequence) can be applied by first contracting those qudits together,
thus forming a single qudit, and then applying the equivalent
single-qudit gate on the combined system.  Swap gates can be used to
rearrange qudits amongst one another when interacting initially
distant qudits.  Finally, one can separate the composite system back
into the original partitions, or as desired, by applying a matrix
decomposition algorithm on an specifically arranged matrix:
\begin{align}
  \begin{bmatrix}
    D_{00}  & D_{01}    & \cdots    & D_{0j} \\
    D_{10}  & D_{11}    & \cdots    & D_{1j} \\
    \vdots  & \vdots    & \ddots    & \vdots \\
    D_{i0}  & D_{i1}    & \cdots    & D_{ij}
  \end{bmatrix}
= \begin{bmatrix}
    A_0     \\
    A_1     \\
    \vdots  \\
    A_i
  \end{bmatrix}
  \begin{bmatrix}
    B_0 & B_1 & \cdots & B_j
  \end{bmatrix}.
\label{eq:decompose tensor}
\end{align}

\noindent
The left-hand side shows the arrangement of the matrices $D_{ab}$,
where $0 \leq a \leq i$ and $0 \leq b \leq j$, of the composite qudit
within a single $m \times n$ matrix, for decomposition into a left
$(i+1)$-dimensional qudit and a right $(j+1)$-dimensional qudit.  A
matrix decomposition algorithm returns the two right-hand side
matrices (or something similar), which have dimensions $m \times k$
and $k \times n$, where $k$ is ideally the \emphdef{rank} of the
left-hand side matrix (the number of linearly independent rows or
columns).  After the decomposition, one can readily extract the
matrices $A_a$ and $B_b$.  Rank-revealing decompositions such as the
\emphdef{singular value decomposition} (SVD) and the
\emphdef{pivoted $QR$ decomposition} (RRQRF), although not strictly
necessary, are beneficial as they minimise $k$ and, hence, the
required storage space.

Note that, due to numerical imprecision, the rank is normally
understood to be the number of eigenvalues above a given threshold
returned by the SVD, and one can introduce a form of approximation by
raising the cut-off value.  MPS was developed as an efficient method
for simulating systems where the number of non-trivial eigenvalues is
limited.  However, for the controlled-$U$ gates in the order-finding
circuit, the eigenvalues are either clearly zero or non-zero, thus
increasing the cut-off value does not provide any benefit (without
invalidating the results).

%% file: chapters/analysis.tex
\section{Order-Finding Circuit Space Analysis}
\label{sec:analysis}

Amplitude formalism state vector simulators are severely limited by
the number of qubits in the circuit.  In the case of the order-finding
circuit, the most resource intensive state under the amplitude
formalism is the state as one applies the final controlled-$U$
operation (\refeq{eq:simplified state}), requiring $O(2^{3l})$ space
for a standard implementation, and $O(2^{2l})$ space for one using
sparse storage.

We now show that the state in \refeq{eq:simplified state} can be
stored much more economically by using the MPS formalism.  We wish to
first divide this state into two qudits: a left-hand side ket which is
the contraction of the upper $2l$ control qubits, and a right-hand
side ket for lower $l$ qubits.  A general state with this division can
be written as
\begin{align}
\ket{\psi}
= \sum_{a=0}^{2^{2l}-1}
  \sum_{b=0}^{2^l-1} A_a B_b \ket{a} \ket{b}.
\label{eq:mps state simple}
\end{align}

\noindent
Here $A_a$ and $B_b$ are complex row and column vectors respectively,
so that the product $A_a B_b$ is a scalar.  We note that in
\refeq{eq:simplified state}, the summation index $i$ spans $r$ values
only, corresponding to the $r$ unique values generated by
$x^i \mod N$.  From this comparison, one concludes that there are only
$r$ non-trivial $B_b$ matrices in \refeq{eq:mps state simple}.  For
linear independence, the $B_b$ matrices must contain at least $r$
rows.  The $A_a$ matrices must possess as many columns as $B_b$ does
rows to satisfy the matrix product.  It follows that for this
particular decomposition, the state can be expressed with
$O(2^{2l} \Paren{1 \times r} + r \Paren{r \times 1})$ space.

One can further decompose the left-hand side ket down to individual
control qubits.  Each of the $2l$ control qubits' two bases can be
represented by matrices no greater than $r \times r$.  This is easily
seen from the structure of the order-finding circuit; the control
qubits only interact with the lower register, and \emph{not} with one
another.  Using this decomposition, \refeq{eq:simplified state} can be
represented using $O(2l \cdot 2 \Paren{r \times r})$ space.

To illustrate the above analysis, see \reftab{tab:mps ranks}.  Here we
have graphed the ranks across the MPS after each controlled-$U$ gate
is applied for the case $N=65$, $x=2$, and $r=12$.  We shall describe
our implementation in detail in \refsec{sec:bench1}.  For now, it
suffices to know that adjacent pairs of numbers define the dimensions
of the intermediate matrix.  One can see that no matrix exceeds
dimensions $r \times r$.

The remainder of the circuit can be completed by first measuring and
discarding the lower register, leaving behind a $2l$ qubit system.
All of the control qubits can then be contracted to a single qudit.
At this point, the MPS representation becomes equivalent to the
amplitude formalism representation, thus it requires $O(2^{2l})$
space.  Therefore, the entire order-finding circuit can be completed
using $O(\max(4lr^2, 2^{2l}))$ space.  Further space reduction is
possible by using sparse matrices---after reading \refsec{sec:bench1},
it should become obvious that the space required before the QFT is
$O(lr)$.

Note that the of value $r$ will arise naturally as the simulation is
run.  It is interesting to observe that in $r < \log_2 N$ cases, one
can efficiently simulate Shor's algorithm by using the semi-classical
QFT \cite{brow07,yora07}.  However, it is exceedingly rare to choose
an $x$ with $r < \log_2 N$ \cite{jozs03}.  Furthermore, for the case
$x = N-1$, it is easy to check that $r=2$, but the algorithm does not
return any non-trivial factors.

\begin{table}
\centering
$\begin{array}{lccccccccc}
\hline
\hline
\multicolumn{10}{c}{\textrm{Increasing Unitary Order}}      \\
\hline
U^{1}   & 1 & \cdots& 1 & 1 & 1 & 1 & 1 & 2 & 1             \\
U^{2}   & 1 & \cdots& 1 & 1 & 1 & 1 & 2 & 4 & 1             \\
U^{4}   & 1 & \cdots& 1 & 1 & 1 & 2 & 4 & 8 & 1             \\
U^{8}   & 1 & \cdots& 1 & 1 & 2 & 4 & 8 &12 & 1             \\
U^{16}  & 1 & \cdots& 1 & 2 & 4 & 8 &12 &12 & 1             \\
U^{8192}& 1 & 2 & 4 & 8 &12 &12 & \cdots&\underline{12} & 1 \\
\hline
\hline
\end{array}$

\caption{%
The ranks across the MPS after completing the order-finding circuit up
to and including the indicated controlled-unitary for $N=65$, $x=2$,
$l=7$.  Increasing time flows down the table.  Adjacent pairs of
numbers define the dimensions of the intermediate matrix.  The lower
register is stored at the rightmost end, so that the second last
number in each row is also equal to the number of non-trivial states
in the lower register.  The underlined value is the order $r=12$.  The
final order of qubits is $q_0, q_1, \cdots, q_{13}$.}
\label{tab:mps ranks}
\vspace{1em}
$\begin{array}{lccccccccc}
\hline
\hline
\multicolumn{10}{c}{\textrm{Decreasing Unitary Order}}      \\
\hline
U^{8192}& 1 & \cdots& 1 & 1 & 1 & 1 & 1 & 2 & 1             \\
U^{4096}& 1 & \cdots& 1 & 1 & 1 & 1 & 2 & 3 & 1             \\
U^{2048}& 1 & \cdots& 1 & 1 & 1 & 2 & 3 & 3 & 1             \\
U^{4}   & 1 & 2 & 3 & 3 & \cdots& 3 & 3 & 3 & 1             \\
U^{2}   & 1 & 2 & 3 & 3 & \cdots& 3 & 3 & 6 & 1             \\
U^{1}   & 1 & 2 & 3 & 3 & \cdots& 3 & 6 &\underline{12} & 1 \\
\hline
\hline
\end{array}$

\caption{%
Applying controlled-unitaries in decreasing powers of two lead to
equal or lower ranks across the MPS (cf. \reftab{tab:mps ranks}),
thus improves both space and time performance.  The final order of
qubits is $q_{13}, q_{12}, \cdots, q_0$.}
\label{tab:mps reverse ranks}
\end{table}

%% file: chapters/bench1.tex
\section{Implementation and Benchmarks}
\label{sec:bench1}

We now describe the details of our simulations and present our
benchmarks.  Preliminary processing steps required before the
order-finding circuit available elsewhere (e.g. see \cite{mike+ike})
and will not be repeated here.  We shall divide the circuit into three
distinct steps:
(1) the application of the controlled-$U$ gates;
(2) the measurement of the lower register; and,
(3) the final quantum Fourier transform.

The circuit consists of $2l$ \emphdef{control qubits}, labelled $q_i$,
$0 \leq i < 2l$, such that the qubit $q_i$ controls the gate
$U^{2^i}$.  The remaining $l$ qubits form the \emphdef{lower register}
$R$.  They are always contracted into a single $2^l$-dimensional qudit
(technically, the circuit also works with a single $N$-dimensional
qudit).  Following our space analysis, we choose only to store the
non-trivial matrices of $R$ (at most $r$), and furthermore, we choose
to always keep it at the rightmost end of the product, so that these
matrices are always column vectors, each containing at most $r$ rows.

The state is initialised in $\ket{0}\ket{1}$.  The $H$ gates may be
applied immediately, or combined into the following controlled-$U$
gates.  We have opted to implement the latter.  In both cases, all of
the matrices representing the state are, at present, simply
scalars---either $0$, $1$, or $1/\sqrt{2}$---and thus the internal
ordering amongst qubits and qudits is negligible.  Note, however, that
for a general entangled state, the ordering is of critical importance.

\subsection{The Controlled-$U$ Gates}

In order to apply the controlled-$U^{2^i}$ gate, we first shift the
control qubit $q_i$ directly to the left of lower register $R$.  Since
$q_i$ is in a product state, or more correctly, its matrices are
currently scalars, one can commute it through the matrix product with
ease, and resize it (i.e. multiplying it by an appropriately sized
square identity matrix) to retain a well-formed MPS.

Now one can contract $q_i$ with $R$, and subsequently apply the
controlled operation.  Let this combined system have matrices denoted
by $D_{ab}$, where $a \in \Brace{0, 1}$ and $b$ are the non-trivial
states of $R$.  Then, in order to separate the system into two
partitions, with $q_i$ on the left and $R$ on the right, one arranges
the matrices $D_{ab}$ into a single matrix as shown on the left-hand
side of \refeq{eq:decompose tensor}.  In practice, instead of
performing each of these steps independently, one can perform the
contraction, apply the controlled-$U$ gate, and arrange the $D_{ab}$
matrices in a single swift blow for efficiency.  Additionally, the $H$
operation on $q_i$, which simply takes the state from $\ket{0}$ to
$\paren{\ket{0} + \ket{1}}/\sqrt{2}$, is also readily combined into
the above steps.

A matrix decomposition algorithm takes the matrix on the left-hand
side of \refeq{eq:decompose tensor} and produces the two matrices on
the right-hand side.  These contain the new matrices associated with
the states of $q_i$ and $R$.  Note that the decomposition is not
unique.  As mentioned earlier, the intermediate dimension is minimised
by using a rank-revealing decomposition, and is at best the rank $k$
of the left-hand side matrix.  However, in this case, we know $k$ is
exactly the number of non-trivial states of $R$.  Furthermore, we have
arranged the matrix such that there are $k$ columns (we can think of
this as pivoting the trivial columns to the end of the matrix).  For
the sake of speed yet still maintaining the space performance of MPS,
whenever one suspects an $m \times n$ matrix $A$ having rank
$k \approx \min(m, n)$, as is the case here, one can use the trivially
simple decomposition:
\begin{align}
A_{m \times n}
= \begin{cases}
    A_{m \times n}~I_{n \times n}, & m \geq n   \\
    I_{m \times m}~A_{m \times n}, & m < n
  \end{cases}.
\label{eq:trivial decomposition}
\end{align}

The above discusses how a single controlled-$U$ gate is implemented.
The circuit consists of $2l$ such gates, which differ only in their
exponents and, therefore, commute.  We find that it is never worse and
often better to apply the $U$s in \emph{decreasing} powers of two.
This is because the sequence generated by $U^{2^i} = x^{2^i} \mod N$,
$i \geq 0$, will eventually lead to a cyclic subsequence.  Applying
the $U$s in increasing powers saturates the rank in the fewest number
of steps, which occurs just before the cycle repeats.  In contrast,
applying the $U$s in decreasing powers leads to performing the same
operations on $R$ at the beginning.  Repeated operations will tend to
populate the same states in $R$ as previous applications, thus leads
to a slower increase in ranks.  Other orderings are also possible, but
we have not considered them due to foresight of the forthcoming QFT.
The rank between the rightmost control qubit and $R$ reaches $r$ once
all unique $U$s have been applied.

\reftab{tab:mps ranks} shows the ranks after each controlled-$U$
operation is applied, for the example $N=64$, $x=2$, and $r=12$.  Time
flows down the table, reflecting applying the $U$s in increasing
powers of two.  Our implementation places the most recently interacted
control qubit furthest to the right, which leads to ordered qubits
upon completing the controlled-$U$ gate sequence:
$q_0, q_1, \cdots, q_{13}$.  Using this ordering, the state can be
stored using $3300$ complex numbers, or approximately $50$KB.
\reftab{tab:mps reverse ranks} shows the same simulation parameters,
now with $U$s applied in decreasing powers, and hence reverse ordered
qubits.  Under this ordering, the same state can be stored using only
$520$ complex numbers, or approximately $8$KB.

\subsection{Measurement of Lower Register}

Measurement of the lower register requires the probability of each
state $i$ in $R$, which can be calculated from first principles using
\refeq{eq:general mps state}.  One can show that this is
\begin{align}
\mathrm{pr}(i)
= \abs{\alpha_{a b \cdots i}}^2
= R_i^{\dag}
  \Brace{\cdots
    \Paren{
      \sum_{b} B_b^{\dag}
      \paren{\sum_a A_a^{\dag} A_a}
      B_b
    }
  \cdots}
  R_i.
\end{align}

It is possible to enforce $\sum_a A_a^\dag A_a = 1$ for all qubits and
qudits in the MPS by using the SVD exclusively.  This allows one to
calculate the probability of a given state of any qubit or qudit by
using its own matrices only.  However, this is not true in our
simulations; we have sacrificed measurement speed for simpler
decompositions, namely \refeq{eq:trivial decomposition}.

Having computed the probabilities, one can apply a projection and
rescale the remaining amplitudes.  Since the projection will reduce
the entanglement in the entire system, one should take this
opportunity to reduce the space consumption.  We apply the RRQRF
pairwise from right to left for this purpose.  We have chosen the
RRQRF only for its rank-revealing property and its speed over the SVD,
and not for any structure in the results.  We have also avoided the
faster LU decomposition as it is considered less numerically stable.

\subsection{The Quantum Fourier Transform}

\begin{figure}
\centering
\includegraphics{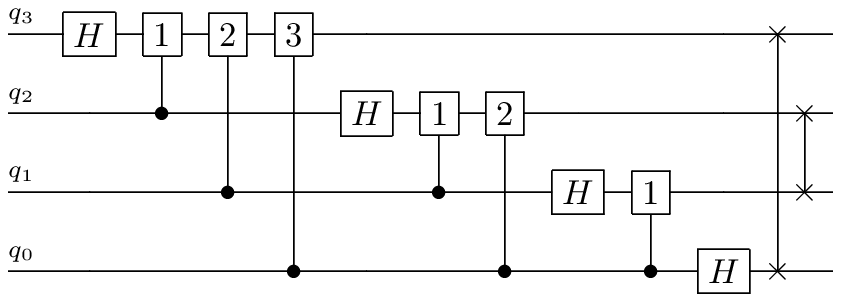}
\caption{%
Standard circuit implementing the quantum Fourier transform on four
qubits.  Controlled-phase gates performing
$\ket{11} \rightarrow \exp\paren{i \pi / 2^n} \ket{11}$ are denoted by
the number $n$ in a square.  Swap gates are denoted by two crosses
joined together by a vertical line.  Long-range gates make this
circuit less than ideal on a MPS simulator.}
\label{fig:qft std}
\vspace{2em}
\includegraphics{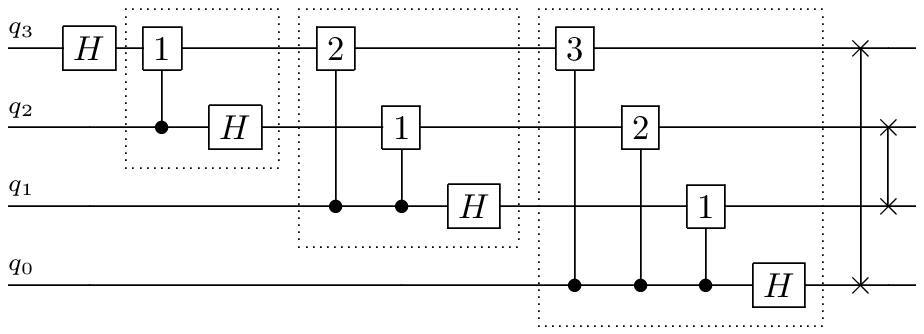}
\caption{%
QFT rearranged to be more compatible with MPS.  Dashed boxes show the
repetitive pattern of the subcircuits.  No matrix decompositions are
required, and all qubits involved are contracted into a single qudit
at the completion of the circuit.}
\label{fig:qft mps}
\vspace{2em}
\includegraphics{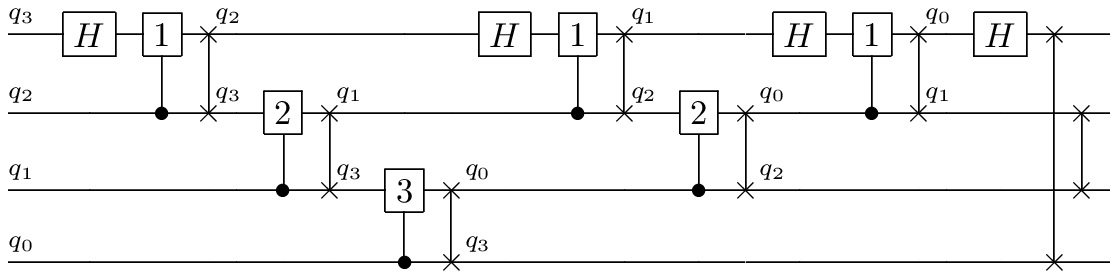}
\caption{%
QFT for a $1$d nearest-neighbour quantum computer is also suitable for
MPS.  Matrix decompositions are only required after every swap
operation.}
\label{fig:qft nn}
\end{figure}

The typical circuit for the QFT is shown in \reffig{fig:qft std}.  The
use of long-range controlled-phase gates make this circuit less than
ideal when using a MPS simulator.  Note that the final swap gates can
be implemented classically and thus do not pose a problem.

To overcome this, one can be rearrange the QFT into $2l$ subcircuits,
each of which includes exactly one more qubit than its predecessor, as
shown in \reffig{fig:qft mps}.  One contracts a single qubit into a
growing qudit at the beginning of each subcircuit, and applies all
gates in the subcircuit directly onto this qudit.  This arrangement is
preferable because one eliminates the need to perform any matrix
decompositions.  Upon completion, a single $2^{2l}$-dimensional qudit
remains (i.e. a $2l$ qubit amplitude formalism state vector).

Another approach is to add swap gates after each controlled-phase
gate, akin to implementing the QFT on a $1$d quantum computer with
only nearest neighbour interactions, as shown in \reffig{fig:qft nn}.
This approach is initially slower, however, it may lead to better
space performance when coupled to the SVD due to truncation of
insignificant eigenvalues.

\subsection{Benchmarks}

\reftab{tab:mps benchmark} shows the CPU times in seconds to simulate
Shor's algorithm for various $N$ and $r$ for each of the described
stages in the algorithm: $t_U$ to apply all controlled-$U$ gates,
$t_{\mathrm{meas}}$ to measure the lower register and minimise the
ranks, and $t_{\mathrm{QFT}}$ to apply the QFT (using
\reffig{fig:qft mps}).  As the memory and time depends heavily on $r$,
$x$ has been chosen ahead of time to maximise $r$ for the given $N$,
and we have also chosen cases where $r \approx N/2$.  All simulations
use a single core and double precision complex numbers; the $l=12$
case would require $1$TB RAM on an amplitude formalism state vector
simulator.  The $l \leq 12$ cases take minutes to run on a laptop
(Intel Core 2 Duo P8400 @ $2.26$GHz, $2$GB RAM), and the $l \geq 13$
cases take hours on a cluster (AMD Opteron @ $2.5$GHz, $32$GB RAM).
For comparison, $l=12$ cases take $970$s (IBM Regatta p$690+$, 512
processes) and $170$s (IBM BlueGene/L, $4096$ processes) on very
highly parallelised amplitude formalism simulators \cite{raed07},
corresponding to days in CPU time.  The same software has been used to
simulate $l=14$ cases, or equivalently $42$ qubits ($64$TB RAM)
\cite{juli10}.

%% file: chapters/table1.tex
\begin{table}
\begin{equation*}
\begin{array}{ccccc}
\hline
\hline
l   & 11                &\quad 12               &\quad 13               &\quad 14               \\
N   & 2033              &\quad 4063             &\quad 8189             &\quad 16351            \\
    & (19 \times 107)   &\quad (17 \times 239)  &\quad (19 \times 431)  &\quad (83 \times 197)  \\
x   & 2                 &\quad 3                &\quad 10               &\quad 2                \\
\hline
r   & 954               &\quad 1904             &\quad 3870             &\quad 8036             \\
t_U & 0.2               &\quad 0.1              &\quad 2.6              &\quad 3.8              \\
t_{\mathrm{meas}}
    & 49                &\quad 76               &\quad 4843             &\quad 12,799           \\
t_{\mathrm{QFT}}
    & 11                &\quad 6.7              &\quad 962              &\quad 1949             \\
t_{\mathrm{total}}
    & 60                &\quad 83               &\quad 5808             &\quad 14,752           \\
\hline
\hline
\end{array}
\end{equation*}

\caption{%
CPU time in seconds the distinct steps of the order-finding circuit
(\reffig{fig:order-finding schematic}):
$t_{U}$ to complete all controlled-$U$s;
$t_{\mathrm{meas}}$ to measure out the lower register; and,
$t_{\mathrm{QFT}}$ perform the QFT.
$t_{\mathrm{total}}$ is the total time.
$x$ is chosen to maximise the order $r$ for the given $N$.
$r$ determines the maximum size of the matrices before the QFT.}
\label{tab:mps benchmark}
\end{table}

%% file: chapters/bench2.tex
\section{Multi-processor Implementation}
\label{sec:bench2}

Simulations of Shor's algorithm benefit immensely from storing the
state as a MPS.  However, single process simulations will ultimately
be memory limited, which is typically resolved by distributed
computing.  We show that MPS is well suited to this regime.  We use
the Message Passing Interface (MPI) \cite{mpiforum} to manage
inter-process communication, the PBLAS library \cite{netlibpblas} for
basic distributed matrix operations, and ScaLAPACK \cite{slug} for
linear algebra.

We assume that our processes are arranged into a 2d grid.  For
simplicity, we shall also assume that the number of processes
$n_{\textrm{proc}}$ is a power of two.  The matrices associated with
the qudits are divided and distributed to these processes.  For our
simulations, we have chosen to divide matrices into $16 \times 16$
blocks, which are distributed cyclically.  It is straightforward to port
the program to this environment; many-qudit operations under MPS are
simply some combination of matrix multiplications, rearranging
or adding sub-matrices, and matrix decomposition.

The simple data distribution described above works well for the
majority of the simulation as the matrices are large and relatively
square (\reftab{tab:mps ranks}).  However, one must take care during
the QFT (\reffig{fig:qft mps}).  This is because the circuit combines
many qubits to form a very high dimensional qudit, with the most
extreme case appearing upon completion, a single $2^{2l}$-dimensional
qudit.  One can easily produce unevenly distributed data, for example,
by creating either $2^{2l}$ $1 \times 1$ matrices (all data stored in
a single process), or a single $2^{2l}$-dimensional row or column
vector (data distributed along a single row or column only).  To
ensure a balanced data distribution throughout the circuit, we set the
matrix's origin (i.e. the process containing the first element) to be
the process whose rank is equal to the first
$\log_2(n_{\mathrm{proc}})$ bits of the state.  This is analogous to
the distribution mechanic of large amplitude formalism simulators,
where the most significant $\log_2(n_{\mathrm{proc}})$ bits of the
state determine the process in which it is stored.

We have further divided the matrices into two aligned matrices, for
the cases where the most significant qubit is $\ket{0}$ and $\ket{1}$.
Since the most significant qubit is the qubit to which the $H$ gate is
applied in each of the subcircuits in \reffig{fig:qft mps}, this
allows $H$ to be applied locally.  Note that, as is the case for the
amplitude formalism, the controlled-phase gates can also be applied
without additional data transfer.

\reftab{tab:mpi benchmark 13} shows the wall-clock times in seconds
for our implementation.  These were run on the same cluster mentioned
in the single process simulations (AMD Opteron @ $2.5$GHz).  The
largest case simulated was $l=15$, using approximately $16$GB RAM
spread over $8$ processes and taking approximately one hour.  Because
these simulations were run on a shared cluster, the execution time can
deviate by $25\%$ or more, depending on the load on the nodes and the
assignment of nodes.  Thus the listed times are the lowest from each
case, obtained from a small number of runs.  The results indicate good
scaling behaviour, doubling the number of processes leads to a
$40$--$60\%$ decrease in the total execution time.

We also mention here that there is another difference between these
simulations and those of \reftab{tab:mps benchmark}.  Recall that
after the measurement projection, we apply a rank-revealing
decomposition across the MPS to reduce the matrix sizes.
Unfortunately, the QR decomposition from ScaLAPACK 1.8.0 (pzgeqpf)
appears to \emph{not} always choose pivots such that the diagonal
(leading) elements of $R$ are in order of descending magnitude, as
described in \cite{slug}.  This appears to be a bug, and the behaviour
means we are unable to determine the numerical rank through this
routine.  Thus in these benchmarks, we have instead opted to use the
SVD (pzgesvd).

%% file: chapters/table2.tex
\begin{table}
\begin{equation*}
\begin{array}{ccccccc}
\hline
\hline
l   & N
    & n_{\mathrm{proc}}
            &~{t_U}~
                    &~{t_{\mathrm{meas}}}~
                                &~{t_{\mathrm{QFT}}}~
                                            &~{t_{\mathrm{total}}}~   \\
\hline
13  & 8189
    & 1     & 3.4   & 6763      & 1539      & 8305      \\
&   & 2     & 2.1   & 3378      & 511       & 3891      \\
&   & 4     & 1.3   & 1636      & 253       & 1890      \\
&   & 8     & 0.7   & 886       & 128       & 1015      \\
&   & 16    & 0.5   & 452       & 76        & 529       \\
\hline
14  & 16,351
    & 1     & 6.3   & 13,173    & 3909      & 17,088    \\
&   & 2     & 3.6   & 5588      & 1430      & 7022      \\
&   & 4     & 1.8   & 2769      & 788       & 3558      \\
&   & 8     & 1.2   & 1471      & 356       & 1829      \\
&   & 16    & 0.7   & 741       & 172       & 914       \\
\hline
15  & 32,663
    & 4     & 5.2   & 7510      & 1934      & 9449      \\
&   & 8     & 3.1   & 3239      & 788       & 4031      \\
\hline
\hline
\end{array}
\end{equation*}

\caption{%
Wall-clock time in seconds for the distinct steps of the order-finding
circuit for three cases:
$l=13$, $N = 8189 = 19 \times 431$, $x=10$, $r=3870$;
$l=14$, $N = 16351 = 83 \times 197$, $x=2$, $r=8036$; and,
$l=15$, $N = 32663 = 89 \times 367$, $x=6$, $r=16104$.
$n_{\mathrm{proc}}$ is the number of MPI processes or CPU cores.}
\label{tab:mpi benchmark 13}
\end{table}